\documentclass[twoside]{llncs}
\usepackage{theorem}  
\usepackage{amssymb}
\usepackage{algorithm,algorithmic}

{\theorembodyfont{\rmfamily} \newtheorem{defi}{Definition}}

\newcommand{\kom}[1]{}

\usepackage{amsmath}  
\usepackage{fancyhdr}           % For LaTeX2.09 you should specify {fancyhdr} 
\usepackage{ifthen}

\newcommand\settitle[2][]{%
 \title{#2}
 \ifthenelse{\equal{#1}{}}%
  {\fancyhead[RO]{\nouppercase #2 \qquad \thepage}}%
  {\fancyhead[RO]{\nouppercase #1 \qquad \thepage}}%
}

\newcommand\setauthors[2]{%
 \author{#2}
  {\fancyhead[LE]{\thepage \qquad \nouppercase #1}}%
}

\fancypagestyle{plain}{%
\fancyhf{}

}

\def\keywordsname{Keywords.}

\begin{document}

\settitle{Quantum Complexity Bounds for Independent Set Problems}

\setauthors{Sebastian D\"orn}{Sebastian D\"orn}

\institute{Institut f\"ur Theoretische Informatik, Universit\"at Ulm, 89069 Ulm,  Germany\\
\tt sebastian.doern@uni-ulm.de}

%Paper id: 25 
%Password: 7e332b 

\maketitle
\thispagestyle{plain}
%---------------------------------------------------------------------------------------------------------------
\begin{abstract}
We present quantum complexity lower and upper bounds for independent set problems in graphs. 
In particular, we give quantum algorithms for computing a maximal and a maximum independent set in a graph. 
We present applications of these algorithms for some graph problems. 
Our results improve the best classical complexity bounds for the corresponding problems.
\end{abstract}
%---------------------------------------------------------------------------------------------------------------
%---------------------------------------------------------------------------------------------------------------
%---------------------------------------------------------------------------------------------------------------
\section{Introduction}
Quantum algorithms have the potential to demonstrate that for some problems quantum computation is more efficiently than classical computation. A goal of quantum computing is to determine whether quantum computers are faster than classical computers. 

The study of the quantum complexity for graph problems is a new area in quantum computing. Two main complexity measures for quantum algorithms have been studied: the quantum query and the quantum time complexity. The quantum query complexity of a graph algorithm $\mathcal A$ is the number of queries to the adjacency matrix or to the adjacency list of the graph made by $\mathcal A$. The quantum time complexity  of an algorithm $\mathcal A$ is the number of basic quantum operations made by $\mathcal A$.

Some new optimal quantum algorithms for graph problems were presented by D\"urr, et al. \cite{DHHM04}. They studied the quantum query complexity for minimum spanning tree, graph connectivity, strong graph connectivity and single source shortest paths  in the adjacency matrix and in the adjacency list model.  Some quantum query lower bounds for graph problems are investigated by Berzina, et al. \cite{BDFLS04} for the dominating set, hamiltonian circuit and the traveling salesman problem. 
Magniez, et al. \cite{MSS05} presented a quantum query algorithm for finding a triangle, and 
Childs and Eisenberg \cite{CE03} for finding a clique of size $k$ in a graph.
Some polynomial time quantum algorithms are given by Ambainis and {\v S}palek \cite{AS06} for computing a maximum matching in a bipartite graph and  for the network flow problem. These algorithms have better running times than the best classical algorithms.

In this paper we study  the potential for speed up of algorithms for independent set problems in graphs with quantum computing.  An independent set is a set of vertices  of a graph in which no two of these vertices are adjacent. A maximal independent set in a graph is an independent set which  is contained in no other independent set. A maximum independent set is a largest independent set of a graph $G$. 

In our paper we present an $O(\sqrt{nm})$ quantum query algorithm for computing a maximal independent set in a graph $G=(V,E)$, where $n=|V|$ and $m=|E|$. We prove that this quantum algorithm is optimal. The quantum time complexity of our algorithm is $O(\sqrt{nm}\log^2 n)$, which is better than the best classical algorithm.

The development of algorithms for the maximum independent set problem is one of the most applicable problem in graph theory. The maximum independent set problem is closely related to the maximum clique and the minimum vertex cover problem. The first exact algorithms for computing a maximum independent set was given by Tarjan and Trojanowski \cite{TT77} with running time of $O(1.2599^n)$. Jian \cite{Jia86} improved the time complexity to $O(1.2346^n)$ and Beigel \cite{Bei99} to $O({1.2227^n})$. Today, the fastest known algorithm was given by Robson \cite{Rob01} with running time of $O(1.1844^n)$. This algorithm is based on a detailed computer generated subcase analysis. 
We construct an $O(1.1488^n)$ quantum time algorithm for computing a maximum independent set. This algorithm is faster than the best classical algorithm by \cite{Rob01}. 

Futhermore, we using the quantum walk clique finding algorithm of Childs and Eisenberg \cite{CE03}, and show that the quantum query complexity of the maximum independent set problem is $O(n^{2\alpha(G)/(\alpha(G)+1)})$, where $\alpha(G)$ is the size of a maximum independent set in $G$. 

Our results are proved using several techniques: Grover search, quantum amplitude amplification and quantum walks.
Maximal und maximum independent set problems have many important applications in graph theory. Our quantum algorithms can  be used as a building blocks for other quantum graph algorithms. 
For example, we give two applications of our independent set algorithms for the following two problems: 
determination of a minimum odd cycle transversal and  finding of a Greedy vertex coloring  of a graph.

The paper is organized as follows: 
In section 2 we give necessary definitions and facts about graph theory and quantum computing.
In section 3 we prove quantum query lower and upper bounds for finding a maximal independent set in a graph. 
In section 4 we develop a quantum time algorithms for finding a maximum independent set. This algorithm is faster than the best classical algorithm for this problem. 
In section 5 we look at the quantum query complexity for computing an independent set of size $k$ in a graph.
At the end of our paper, we present some graph applications of our quantum independent set algorithms.

%--------------------------------------------------------------------------------------------------------------
%--------------------------------------------------------------------------------------------------------------
\section{Preliminaries}
\subsection{Graph Theory}
Let $G=(V,E)$ be a undirected graph, with $V=V(G)$ and $E=E(G)$ we denote the set of vertices and edges of $G$. Let $n=|V|$ be the number of vertices and $m=|E|$ the number of edges of $G$. 
We denote with $N_G(v)$ the set of all adjacent vertices to $v\in V$  and $d_G(v):=|N_G(v)|$.  
Let $\Delta(G):= \max \{ d_G(v)\ |\ v\in V(G)\}$ be the \textit{maximum degree} of $G$. 
The graph $G_{-S}$ is obtained from $G$ by deleting the vertices $S \subset V$ and the incident edges.
We denote with $[n]$ the set $\{1, \ldots, n\}$.

\begin{defi}
A set of vertices $V'\subseteq V$ is called \textit{independent}, if for all distinct vertices $u,v \in V'$ it holds $\{u,v\}\notin E(G)$ .
The set $V'$ is called \textit{maximal}, if there is no independent set $V''\subseteq V$ with $V' \subset V''$.
 A \textit{maximum independent set} is a largest independent set of $G$.
	By $\alpha(G)$ we denote the \textit{independence number} of $G$, i.e. the size of a maximum independent set in $G$.
\end{defi}
We consider the following models for accessing information in undirected graphs:
  
  \begin{description} 
  \item[Adjacency matrix model:] 
  		Given is the adjacency matrix $A\in\{0,1\}^{ n\times n}$ of $G$
  		with $A_{i,j}=1$ iff $\{i,j\} \in E$.

  \item[Adjacency list model:] 
  		Given are the degrees $d_G(1),\ldots, d_G(n)$ of the vertices and for every $i \in V$ an
    	array with its neighbours $f_i : [d_G(i)] \rightarrow [n]$.  
    	The value $f_i(j)$ is the $j$-th neighbour of $i$. 
  \end{description}
In the following, we denote by {\bf M} and {\bf L} the input model of the graph as  
adjacency matrix ({\bf M}) and as adjacency list ({\bf L}). 
%---------------------------------------------------------------------------------------------------------------
%---------------------------------------------------------------------------------------------------------------
\subsection{Quantum Computing}  
For the basic notation on quantum computing, we refer the reader to the textbook by Nielsen and Chuang \cite{NC03}.
For the quantum algorithms included in this paper we use the following two complexity measures:

  \begin{description} 
  		\item[Quantum Query Complexity:]
					The quantum query complexity of a graph algorithm $\mathcal A$ is the number of queries to the 
					adjacency matrix or to the adjacency list of the input graph made by $\mathcal A$.
	 		\item[Quantum Time Complexity:]
	 				The quantum time complexity  of a graph algorithm $\mathcal A$ is the number of basic quantum
	 				operations made by $\mathcal A$.
  \end{description}

 In this paper, we use the following special case of the Ambainis method \cite{Amb02} to prove lower bounds for the quantum query complexity.

\begin{theorem}{\rm \cite{Amb02}}\label{Amb02}
Let $A \subset \{0, 1\}^n, B\subset \{0, 1\}^n$ and $f:\{0,1\}^n \rightarrow \{0,1\}$ 
such that $f(x)=1$ for all $x\in A$, and $f(y)=0$ for all $y \in B$. 
Let~$m$ and~$m'$ be numbers such that
\begin{enumerate}
\item 
for every $(x_1,\ldots,x_n)\in A$ there are at least $m$ values $i \in \{1,\ldots, n\}$ 
such that $(x_1, \ldots , x_{i-1}, 1 - x_i, x_{i+1}, \ldots , x_n)\in B$,
\item 
for every $(x_1,\ldots,x_n)\in B$ there are at least $m'$ values $i \in \{1,\ldots, n\}$ 
such that $(x_1, \ldots, x_{i-1}, 1 - x_i, x_{i+1}, \ldots , x_n)\in A$.
\end{enumerate}
Then every bounded-error quantum algorithm that computes $f$ has quantum query complexity
$\Omega(\sqrt{m \cdot m'})$.
\end{theorem}

Now we give two tools for the construction of our quantum algorithms.

\paragraph{Quantum Search.}
A search problem is a subset $P \subseteq [N]$ of the search space~$[N]$.
With~$P$ we associate its characteristic function $f_P: [N] \rightarrow \{0,1\}$ with
\[
f_P(x)=
\begin{cases}
1, & \text{if } x \in P,\\
0, & \text{otherwise}.
\end{cases}
\]%$f_P(x)= 1$ if $x \in P$, and 0 otherwise.
Any $x \in P$ is called a solution to the search problem.
Let $k = |P|$ be the number of solutions of~$P$.
\begin{theorem}{\rm \cite{Gro96,BBHT98}}\label{Grover}
For $k > 0$,
the expected quantum query complexity for finding one solution of $P$ is $O(\sqrt{N/k})$, 
and for finding all solutions, it is $O(\sqrt{kN})$. 
Futhermore, 
whether $k>0$ can be decided  in $O(\sqrt{N})$ quantum queries to $f_P$. 
\end{theorem}
We denote with {\scshape All Quantum Search}[$f_P$] an application of Grover search algorithm that 
computes the set  of all solutions.

\paragraph{Amplitude Amplification.}
Let $\mathcal A$ be an algorithm for a problem with small success probability at least $\epsilon$.
Classically, we need $\Theta(1/\epsilon)$ repetitions of $\mathcal A$ to increase 
its success probability from $\epsilon$ to a constant, for example 2/3.  
The corresponding technique in the quantum case
is called amplitude amplification.

\begin{theorem}{\rm \cite{BHMT00}}\index{Amplitude amplification}
Let $\mathcal A$ be a quantum algorithm with one-sided error and success probability at
least $\epsilon$. 
Then there is a quantum algorithm $\mathcal B$ that solves $\mathcal A$ with success
probability 2/3 by $O(\frac{1}{\sqrt{\epsilon}})$ invocations of $\mathcal A$.
\end{theorem}

\begin{remark}\label{log}
Our quantum algorithms output an incorrect answer with a constant probability $p$.
If we want to reduce the error probability to less than $\epsilon$, we repeat each quantum subroutine $l$ times, where $p^{ l} \leq \epsilon$.
It follows, that we have to repeat each quantum subroutine $l = O(\log n)$ times, to make the probability of a correct answer greater than $1-1/n$. This increases the running time of all our algorithms by a logarithmic
factor. Furthermore, the running time of Grover search is bigger that its query complexity by another logarithmic factor. 
\end{remark}

%----------------------------------------------------------------------------------------

%--------------------------------------------------------------------------------------------------------------
%--------------------------------------------------------------------------------------------------------------
%--------------------------------------------------------------------------------------------------------------
\section{Maximal Independent Set}
In this section we study the quantum query complexity of the following problem:\\[2ex]
{\bf Maximal Independent Set}\index{Independent Set!maximal}: Given a graph $G=(V,E)$, compute a maximal independent set in $G$.

\smallskip
 
We present an $O(\sqrt{nm})$ quantum query algorithm for computing a maximal independent set in a graph.
Then we show that this algorithm is optimal in the adjacency model, by proving a lower bound of $\Omega(n^{1.5})$.
Let $G=(V,E)$ be a graph and $v \in V$. For the application of quantum search, we define a search function $f_{G,v}: V \rightarrow \{0,1\}$ with $f_{G,v}(x)=1$ if $x \in N_G(v)$, and zero otherwise.
%----------------------------------------------------------------------------------------
\begin{algorithm}[H]%[h!tp]
{\bf Input: } Graph $G=(V,E)$.  \\
{\bf Output: } Maximal independent set $V'$.  \\
{\bf Complexity: } {\bf M:} $O(n^{1.5})$, {\bf L:} $O(\sqrt{nm})$ quantum queries.

\begin{algorithmic}[1]
	\STATE $V':=\emptyset$, $F:=G$
	\WHILE {$V(F)\neq \emptyset$}
		\STATE Choose $v \in V(F)$ 
		\STATE $V':= V' \cup \{v\}$
		\STATE $W:= ${\scshape All Quantum Search}$[f_{F,v}] \cup \{v\}$
		\STATE $F := F_{-W}$
	\ENDWHILE
\end{algorithmic}
\caption{\,{\scshape Maximal Independent Set}}
\end{algorithm}
%----------------------------------------------------------------------------------------

\begin{theorem}
 The quantum query complexity of the {\scshape Maximal Independent Set} algorithm is $O(n^{1.5})$ in the adjacency matrix  model and $O(\sqrt{nm})$ in the adjacency list model. 
\end{theorem}
\begin{proof}
It is clear, that the algorithm computes a maximal independent set, but not necessarily a maximum independent set.
We use Grover's search to find the set $W$ of all neighbours of the vertex $v$. Then we delete all vertices of $W$ from the graph $F$. Every vertex is deleted at most once. 
In the adjacency matrix model, every vertex is found in $O(\sqrt n)$ quantum queries to the adjacency matrix and total we use $O(n^{1.5})$ quantum queries in the adjacency matrix model. 

In the adjacency list model, processing a vertex $v$ costs $O(\sqrt{d_G(v) a_v})$ quantum queries,
where $a_v$ is the number of vertices in $F$ which are adjacent to $v$.
Since $\sum_v a_v \le n$, then the quantum query complexity is upper-bounded by the Cauchy-Schwarz inequality:
\[
			\sum_v \sqrt{d_G(v) a_v} \leq \sqrt{\sum_v d_G(v)} \sqrt{\sum_v a_v}  = O(\sqrt{m n}).  \ \ \rule{1ex}{1ex}
\]
\end{proof}
In order to get the success probability of $1-1/n$, we need to amplify the success probability of each subroutine by repeating it $O(\log n)$ times, see Remark \ref{log}. Therefore we get:

\begin{corollary}
 The quantum time complexity of the {\scshape Maximal Independent Set} algorithm is $O(n^{1.5}\log^2 n)$ in the adjacency matrix model and $O(\sqrt{nm}\log^2 n)$ in the adjacency list model.
\end{corollary}

Now we prove a $\Omega(n^{1.5})$ quantum query lower bound for the maximal independent set problem with the method of Ambainis \cite{Amb02} and analog to Berzina et al. \cite{BDFLS04}. 
Consequently the  {\scshape Maximal Independent Set} quantum algorithm is optimal in adjacency matrix model.

\begin{theorem}\label{LB-MAIS}
The maximal independent set problem requires $\Omega(n^{1.5})$ quantum queries to the adjacency matrix.
\end{theorem}
\begin{proof}
We construct the sets $A$ and $B$ for the usage of Theorem \ref{Amb02}. 
Let $f$ be the Boolean function which is one, iff there is a maximal independent set of size $2n$. 
%Let $f$ be the Boolean function, which is one if there is a maximal independent set vertices.
The set $A$ consists of all graphs $G=(V,E)$ with $|V|=3n+1$ satisfying the following requirements:
1. There are $n$ mutually not connected red vertices. 
2. There are $2n$ green vertices not connected with the red ones. Green vertices are grouped in pairs and each pair is connected by edge.
3. There is a black vertex  which is connected to all red and green vertices.
Let $V'$ be the set of $n$ red vertices and one green vertex of each pairs. 
%This set $V'$ is the same set of vertices in $A$ and $B$.
Then $V'$ is a maximal independent set in $G$.
The value of the function $f$ for all graphs $G\in A$ is 1.

The set $B$ consists of all graphs $G'=(V,E)$ with $|V|=3n+1$ satisfying the following requirements:
1. There are $n+2$ mutually not connected red vertices.
2. There are $2n-2$ green vertices not connected with red ones, green vertices are
	grouped in pairs and each pair is connected by edge.
3. There is a black vertex  which is connected to all red and green vertices.
%Let $V'$ be the same set of vertices as above, i.e. it consists now of $n+1$ red vertices and $n-1$ green vertices. Then $V'$ is not a maximal independent set in $G'$. 
The value of the function $f$ for all graphs $G'\in B$ is 0, since there no maximal independent of size $2n$ in $G'$. 

From each graph $G \in A$, we can obtain $G'\in B$ by deleting one edge between two green vertices, then $l = n = O(n)$.  From each graph $G'\in B$, we can obtain $G\in A$ by adding an edge between two red vertices, then $l' = (n+2)(n+1)/2 = O(n^2)$.
By Theorem \ref{Amb02}, the quantum query complexity of the maximal independent set problem is $\Omega(\sqrt{l \cdot l'}) = \Omega(n^{1.5})$.\ \ \rule{1ex}{1ex}
\end{proof}

%--------------------------------------------------------------------------------------------------------------
%--------------------------------------------------------------------------------------------------------------
%--------------------------------------------------------------------------------------------------------------
\section{Maximum Independent Set}
Now we are interested in the quantum time complexity for computing a largest independent set in a graph.
This is a well known {\bf NP}-hard problem, which is important for many other applications in computer science and graph theory. \\[2ex]
 {\bf Maximum Independent Set}: Given a graph $G=(V,E)$, compute an independent set $V'\subseteq V$ with $|V'|=\alpha(G)$.

\smallskip

The first exact algorithms for the maximum independent set problem is given by Tarjan and Trojanowski \cite{TT77} with running time of $O(1.2599^n)$. Jian \cite{Jia86} improved the time complexity to $O(1.2346^n)$, Beigel \cite{Bei99} to $O({1.2227^n})$, and Robson \cite{Rob01} to $O(1.1844^n)$.
%Beigel [Bei99] presented another algorithm with time complexity of $O({1.2227^n})$. 
%Today, the fastest algorithm was given by Robson [Rob01] with running time of $O(1.1844^n)$. 
The algorithm by Robson is today the fastest algorithms, it based on a detailed computer generated subcase analysis (number of subcases is in the tens of thousands).  
%It is an open problem in classical computing (see [Woe03]) to construct an exact algorithm for the maximum independent set problem with time complexity $O(c^n)$ for some $c<1.1$.
%We show how to improve the upper complexity bounds for the maximum independent set problem. A naive quantum search about all subsets of the vertices gives a search space of $2^n$ elements and a quantum time algorithm with running time $O(1.4143^n)$. 
We construct a quantum algorithm which is faster than the best classical algorithm. 
%We speed up the solution of finding a maximum independent set in a graph, and 
Our quantum algorithm has running time of $O(1.1488^n)$. This is no query algorithm, in this algorithm we count the time steps to compute a maximum independent set. Our algorithm combines a classical probabilistic algorithm with the quantum amplitude amplification.
First we need two simple facts from maximal independent set theory.

\begin{lemma}\label{MISL1}
For a path $P_n$ and a cycle $C_n$ with $n$ vertices, it holds
\begin{align*}
		\alpha(P_n) = \left\lceil \frac{n}{2} \right\rceil \text{ and } \alpha(C_n) = \left\lfloor  \frac{n}{2}\right\rfloor.
\end{align*}
\end{lemma}
\begin{lemma}\label{MISL2}
Let $G$ be a simple graph with $\Delta(G) \leq 2$. Then all the components of $G$ are paths and cycles.
\end{lemma}

With the application of the above two Lemmas we can construct a quantum time algorithm for the  maximum independent set problem. If the maximum degree of the graph is at most two, we denote with ${\rm Paths}(G)$ and ${\rm Cycles}(G)$  the set of all paths and cycles in the graph $G$. The computation of the maximum independent set of a path or a cycle is then a simple task. We denote with ${\rm MIS}(G')$ the  maximum independent set in a graph $G'$, which is a path or a cycle.
%----------------------------------------------------------------------------------------
\begin{algorithm}[H]%[h!tp]
{\bf Input: } A graph $G=(V,E)$.  \\
{\bf Output: } Maximum independent set (MIS) $V'$.  \\
{\bf Complexity: } {\bf M, L:} $O(1.1488^n)$ quantum steps.

\begin{algorithmic}[1]
		\STATE $F:=G$, $V':=\emptyset$
		\WHILE {$V(F) \neq \emptyset$}
		\IF{ $\Delta(F) \leq 2$}
				\STATE 	$V' := \bigcup_{P \in {\rm Paths}(F)} {\rm MIS}(P) \cup 
									 \bigcup_{C \in {\rm Cycles}(F)} {\rm MIS}(C)$
				\STATE {\bf return}[$V'$]
		\ENDIF
		\STATE Find $v \in V(F)$ with $\Delta(F) = {\rm deg}_F(v)$
		\STATE $a\in_R \{0,1\}$ %{\bf random}$[\{0,1\}]$
		\IF{$a=0$}
			\STATE $F := F_{-\{v\}}$
		\ELSE
			\STATE $V':= V' \cup \{v\}$
			\STATE $F:= F_{-N_{F}[v]}$
		\ENDIF
		\ENDWHILE
		\STATE Apply {\scshape Amplitude Amplification}
\end{algorithmic}
\caption{\,{\scshape Maximum Independent Set}}
\end{algorithm}
%----------------------------------------------------------------------------------------
\begin{theorem}
The expected quantum time complexity of the {\scshape Maximum Independent Set} algorithm is $O(2^{n/5}) = O(1.1488^n)$. 
\end{theorem}
\begin{proof}
The {\scshape Maximum Independent Set} algorithm combines a classical probabilistic algorithm with the quantum amplitude amplification \cite{BHMT00}. We show that the probability for computing a maximum independent set with	the classical algorithm is at least $\epsilon=(1/2)^{2n/5}$. 
To obtain a quantum algorithm, we just use quantum amplitude amplification like \cite{Amb05}. We search for a largest independent set, which can be model by the maximum finding algorithm by D\"urr and H\o yer \cite{DH96}. Then we increase the success probability to a constant, by repeating the algorithm $O(\frac{1}{\sqrt{\epsilon}}) = O(2^{n/5})$ times. Considering this, we obtain the  indicated quantum time complexity.

Now we prove that the probability for computing a maximum independent set with	the classical algorithm is at least $\epsilon=(1/2)^{2n/5}$.
In the first steps of this algorithm, we check if the maximal degree of the graph is smaller or equal than two. If this is true,  we apply Lemma \ref{MISL1} and Lemma \ref{MISL2}, and compute the maximal independent set $V'$.
Otherwise we choose a vertex $v$ with maximal degree, and a random variable $a \in \{0,1\}$.
%We compute the probability that the algorithm without the loop finds a maximum independent set.
If $a=0$, we assume that $v$ is not in the maximum independent set $V'$, and then we delete the vertex $v$ from $F$.
In the other case, the vertex $v$ is in the maximum independent set $V'$. We delete $v$ and the set of all neighbours $N_F(v)$ from $F$. Since $\Delta(G) \geq 3$, we delete at least four vertices. 

The task is now to determine how many expected number of steps $x$ must we do if $F$ is empty. We choose the value of $a$ with uniform distribution from $\{0,1\}$. If $a=0$ we delete one vertex and if $a=1$ we delete at least four vertices of $F$, such that $n \geq \frac{1}{2} \left ( 1x + 4x\right)$.
Then it is $x \leq 2n/5$ and 
\[
	{\rm Prob}(V' \text{ is a MIS}) \geq (1/2)^{2n/5}.
\]
%\begin{align*}
%		{\rm Prob}(V' \text{ is a MIS}) 
%				\geq \left(\frac{1}{2}\right)^{2n/5}.
%\end{align*}
Now we apply the amplitude amplification, and repeat the procedure 
\[
			O(1/\sqrt{{\rm Prob}(V' \text{ is a MIS})}) = O(2^{n/5}) = O(1.1488^n)
\]
times, to compute a maximum independent set $V'$ of $G$. \ \ \rule{1ex}{1ex}
\end{proof}

\begin{theorem}\label{LB-MIS}
The maximum independent set problem requires $\Omega(n^{1.5})$ quantum queries to the adjacency matrix.
\end{theorem}
\begin{proof}
Every maximum independent set is a maximal independent set, and this requires $\Omega(n^{1.5})$ quantum queries to the adjacency matrix. \ \ \rule{1ex}{1ex}
\end{proof}
%---------------------------------------------------------------------------------------------------------------
%---------------------------------------------------------------------------------------------------------------
%---------------------------------------------------------------------------------------------------------------
\section{Independent Set of size $k$}
In this section, we regard the quantum query complexity of the following problem:\\[2ex]
{\bf $\mathbf{k}$-Independent Set:} Given a graph $G$ and an integer $k$, compute an independent set of size $k$ (if there is one).

\smallskip

We use clique finding for proving the quantum query complexity of the $k$-independent set problem. 
A clique of size $k$ in $G$ is a complete subgraph with $k$ vertices. The quantum query complexity for finding such a clique is $\widetilde{O}(n^{(5k-2)/(2k+4)})$ for $k\leq 5$ and $O(n^{2k/(k+1)})$ for $k\geq 6$, see Childs and Eisenberg \cite{CE03}. The corresponding quantum algorithm for this problem uses quantum walks \cite{AAKV01}. Quantum walks are the quantum counterpart of random walks, this is a recent technique for the construction of new quantum algorithms (see [Amb03, Amb04, Sze04, MSS05, MN05, BS06]). Ambainis \cite{Amb04} constructed a fundamental quantum walk algorithm for the element distinctness problem. The authors of \cite{CE03} apply the Ambainis quantum walk for searching a clique of size $k$ in a graph. From the analysis of this $k$-clique algorithm we get immediately the following theorem:

\begin{theorem}
The quantum query complexity of the $k$-independent set problem is $O(n^{(5k-2)/(2k+4)})$ for $k\leq 5$ and $O(n^{2k/(k+1)})$ for $k\geq 6$.
\end{theorem}
\begin{corollary}
The quantum query complexity of the maximum independent set problem is $O(n^{2\alpha(G)/(\alpha(G)+1)})$.
\end{corollary}

\begin{theorem}
The $k$-independent set problem requires $\Omega(n)$ quantum queries.
\end{theorem}

\begin{proof}
The proof is a reduction from $k$-clique. Let $G=(V,E)$ be a graph, then a clique of size $k$ in $G$ is an  independent set of size $k$ in $G'=(V,E')$ with $E' = \{\{u,v\}\in V \times V\ |\ \{u,v\} \notin E)$. Finding a $k$-clique requires $\Omega(n)$ quantum queries (see [BDHHMSW01]). Therefore we obtain the  indicated quantum query lower bound.\ \ \rule{1ex}{1ex}
\end{proof}
%--------------------------------------------------------------------------------------------------------------
%--------------------------------------------------------------------------------------------------------------
%--------------------------------------------------------------------------------------------------------------
\section{Graph Application}
\subsection{Minimum Odd Cycle Transversal}
\begin{definition}
		Let $G=(V,E)$ be a graph, an \textit{odd cycle transversal} of $G$ is a subset of vertices whose deletion makes the graph bipartite. The size of a minimum odd cycle transversal is called  the \textit{vertex bipartization number}.
\end{definition}
\textbf{Minimum Odd Cycle Transversal}: Given a graph $G$,  compute a minimum odd cycle transversal of $G$.

\smallskip

We use our maximum independent set algorithm and a decomposition theorem of Raman and Saurabh [RS05] for finding a minimum odd cycle transversal with quantum time complexity of $O(1.5819^n)$. This improves the best classical known time complexity bound of $O(1.62^n)$ by \cite{RS05}. %$O(1.7724^n)$ by Byskov and  
First we give the  decomposition theorem of \cite{RS05}, and then we present the corresponding quantum algorithm.

\begin{theorem}\rm{\cite{RS05}}\label{MOCT}
		Let $G=(V,E)$ be a connected graph and let $O$ a minimum odd cycle transversal. Then $V\backslash O$ can be decomposed as $V_1$ and $V_2$ such that $V_1$ is a maximal independent set of $G$ and $V_2$ is a maximum independent set of $G_{-V_1} $.
\end{theorem}
The correctness of the following {\scshape Minimum Odd Cycle Transversal} algorithm follows immediately from Theorem \ref{MOCT}.
%----------------------------------------------------------------------------------------
\begin{algorithm}[H]%[h!tp]
{\bf Input: } Connected graph $G=(V,E)$.  \\
{\bf Output: }Minimum Odd Cycle Transversal $O$.  \\
{\bf Complexity: } {\bf M, L:}  $O(1.5819^n)$ quantum steps.

\begin{algorithmic}[1]
			\STATE $O:=V$
			\STATE $\mathcal I:=$ Set of all maximal independent sets of $G$
			\FOR{$V_1 \in \mathcal I$}
					\STATE $V_2:=${\scshape Maximum Independent Set}$[G_{-V_1}]$
					\IF {$|O|>|V\backslash (V_1\cup V_2)|$}
							\STATE $O:=V\backslash (V_1\cup V_2)$
					\ENDIF
			\ENDFOR
\end{algorithmic}
\caption{\,{\scshape Minimum Odd Cycle Transversal}}
\end{algorithm}
%----------------------------------------------------------------------------------------
We use our quantum maximum independent set quantum algorithm for computing a minimum odd cycle transversal of the graph $G$. Before we prove the quantum time complexity of the algorithm, we need the number of maximal independent sets in a graph, which are smaller than a constant.

\begin{theorem}\rm{\cite{Epp03}}
		Let $G$ be a graph and $k$ be a constant. The number $M(k)$ of maximal independent sets $I$ for which $|I|\leq k$ is at most 
		\begin{align*}
				M(k)= \left\{ \begin{array}{ll}
							3^{4k-n}4^{n-3k}, & k \leq \left\lfloor \frac{n}{3} \right\rfloor \\
							3^{n/3},					& k \geq \left\lfloor \frac{n}{3} \right\rfloor + 1.
							\end{array} \right.	
		\end{align*}
		Furthermore, there is an algorithm for listing all maximal independent sets of size at most $k$ in time $O(3^{4k-n}4^{n-3k})$ and $O(3^{n/3})$. 
\end{theorem}

\begin{theorem}
The quantum time complexity of the {\scshape Minimum Odd Cycle Transversal} algorithm is $O(1.5819^n)$.
\end{theorem}
\begin{proof}
The time complexity of the algorithm is upper bounded by
\begin{align*}
		\sum_{k=1}^n M(k) \cdot 2^{(n-k)/5} 
					&= \sum_{k=1}^{\left\lfloor n/3 \right\rfloor} 3^{4k-n}4^{n-3k} \cdot 2^{(n-k)/5} +
					  \sum_{k=\left\lfloor n/3 \right\rfloor+1}^{n} 3^{n/3} \cdot 2^{(n-k)/5} \\
					&= \left(\frac{4 \cdot 2^{1/5}}{3} \right)^n \cdot
							\sum_{k=1}^{\left\lfloor n/3 \right\rfloor} \left(\frac{3^4}{4^3 \cdot 2^{1/5}}\right)^k
							+ \left(3^{1/3}\right)^n \sum_{k=0}^{\left\lfloor 2n/3 \right\rfloor} 2^{k/5}
\end{align*}
\begin{align*}
					&\leq O\left(\left(\frac{4 \cdot 2^{1/5}}{3} \right)^n \cdot  
							\left(\frac{3^4}{4^3 \cdot 2^{1/5}}\right)^{n/3}\right) + 
							O\left(\left(3^{1/3}\right)^n  \cdot 2^{2n/3 \cdot 1/5}\right) \\
					&=O(1.5819^n) + O(1.5819^n).\ \ \rule{1ex}{1ex}
\end{align*}
\end{proof}
%--------------------------------------------------------------------------------------------------------------
%--------------------------------------------------------------------------------------------------------------
\subsection{Coloring}
Given is a graph $G$, a coloring of $G$ is an assignment of the vertices, such that the endpoints of each edge are assigned two different colors. Every color class is a vertex set without induced edges, such a vertex set is an independent set. We consider	the following problem:\\[2ex]
\textbf{Vertex-Coloring: } Given a graph $G=(V,E)$, compute a vertex coloring of $G$ (let $k$ be the number of different colors).

\smallskip

We can apply our quantum maximal independent set algorithm to compute a vertex-coloring.
We determine a maximal independent set $W$ of the graph $G$. We assign all vertices of $W$ with color $i$ (at the beginning $i=1$).  Then we delete all the vertices of $W$ from $G$ and increase $i$. We repeat this procedure as long as there are vertices in $G$. The result of this procedure is a coloring of $G$ with $k$ colors, which is not necessary the smallest number for coloring of $G$.

\begin{theorem}
 The quantum time complexity of the vertex-coloring problem is $O(k n^{1.5}\log^2 n)$ in the adjacency matrix model and 
 $O(k \sqrt{nm}\log^2 n)$ in the adjacency list model. 
\end{theorem}
%--------------------------------------------------------------------------------------------------------------
%--------------------------------------------------------------------------------------------------------------
%--------------------------------------------------------------------------------------------------------------
\section*{Conclusion}
We give a summary of the quantum complexity for the regarded independent set problems:

%-----------------------------------------------------------------------------------------------------------
\begin{center}
	\begin{tabular}{|l|l|l|l|}
	\hline
	{\bf Problem}		&   {\bf Quantum Query} &   {\bf Quantum Query}  & {\bf Quantum Time} \\ 
									&   {\bf  Lower Bound} 	&   {\bf  Upper Bound}   & {\bf Complexity} \\ \hline 
	Maximal 				& {\bf M:} $\Omega(n^{1.5})$ &  {\bf M:} $O(n^{1.5})$ &	{\bf M:} $O(n^{1.5}\log^2 n)$		\\
	Independent Set & 													& 	{\bf L:} $\ O(\sqrt{nm})$	 & 	{\bf L:} $\ O(\sqrt{nm}\log^2 n)$																\\ \hline																			

	Maximum 		& {\bf M:} $\Omega(n^{1.5})$ &  {\bf M:} $O(n^{2\alpha(G)/(\alpha(G)+1)})$ &	{\bf M:} $O(1.1488^n)$		\\
	Independent Set & 													&  & 		\\ \hline																			

	Independent Set		 & {\bf M:} $\Omega(n)$ &  		{\bf M:} $O(n^{(5k-2)/(2k+4)})$		& {\bf M:} $O(1.1488^n)$ \\
	of size $k$				 & 											&   	\hspace{0.5cm} $O(n^{2k/(k+1)})$			 & 	\\ \hline											
	
	%Minimum Odd				& {\bf M:} $$ &  {\bf M:} 	&		{\bf M:} $O(1.5819^n)$		\\
	%Cycle Transversal & 													& 	{\bf L:}  & 	{\bf L:} $\ O(1.5819^n)$	\\ \hline																				
	\end{tabular}
\end{center}
There are some interesting open questions in the area of independent set problems. It is an open problem in classical computing to construct an exact algorithm for the maximum independent set problem with time complexity $O(c^n)$ for some $c<1.1$. 
Can we solve this problem with a quantum algorithm? Another interesting task is to improve the $\Omega(n)$ quantum query lower bound for the $k$-independent set problem. The improvement of this lower bound would implies a better lower bound of the $k$-clique problem too.

%---------------------------------------------------------------------------------------------------------------
%---------------------------------------------------------------------------------------------------------------
%---------------------------------------------------------------------------------------------------------------
%\section*{Acknowledgments}
%For helpful comments I am grateful to Jacobo Tor\'an.
%---------------------------------------------------------------------------------------------------------------
%---------------------------------------------------------------------------------------------------------------
%---------------------------------------------------------------------------------------------------------------

\end{document}